# A model for the emergence of the genetic code as a transition in a noisy information channel

Tsvi Tlusty

*Department of Physics of Complex Systems,*
*Weizmann Institute of Science, Rehovot, Israel, 76100,*
*E-mail: tsvi.tlusty@weizmann.ac.il*

**Abstract**. The genetic code maps the sixty-four nucleotide triplets (codons) to twenty amino-acids. Some argue that the specific form of the code with its twenty amino-acids might be a 'frozen accident' because of the overwhelming effects of any further change. Others see it as a consequence of primordial biochemical pathways and their evolution. Here we examine a scenario in which evolution drives the emergence of a genetic code by selecting for an amino-acid map that minimizes the impact of errors. We treat the stochastic mapping of codons to amino-acids as a noisy information channel with a natural fitness measure. Organisms compete by the fitness of their codes and, as a result, a genetic code emerges at a supercritical transition in the noisy channel, when the mapping of codons to amino-acids becomes nonrandom. At the phase transition, a small expansion is valid and the emergent code is governed by smooth modes of the Laplacian of errors. These modes are in turn governed by the topology of the error-graph, in which codons are connected if they are likely to be confused. This topology sets an upper bound – which is related to the classical map-coloring problem – on the number of possible amino-acids. The suggested scenario is generic and may describe a mechanism for the formation of other error-prone biological codes, such as the recognition of DNA sites by proteins in the transcription regulatory network.

1. Introduction

Genes are molecular words written in a language of four letters, the nucleic bases U, C, G and A. Proteins are written in a language of twenty letters, the amino-acids. All organisms read genes and translate them into proteins. The translation of genes to proteins uses a molecular dictionary called the genetic code (Fig 1A). In the code, each amino-acid is encoded by specific triplets of bases known as codons. There are 64 possible codons, but not all of them can be actually discerned by the translation apparatus (Crick, 1966; Osawa et al., 1992) and the effective number of codons is therefore estimated to be somewhere between 48 and 64. Since there are at least 48 discernable codons and only 20 amino-acids (and 3 stop codons), the code is degenerate in the sense that several codons represent the same amino-acid.

The genetic code exhibits significant order (Woese, 1965; Haig and Hurst, 1991): The codons for a given amino-acids are grouped into contiguous domains in the table (Fig.1A). Furthermore, codons that are close in the table tend to encode chemically similar amino-acids. Several studies tried to answer the fundamental question of how such a map with its well-ordered association of codons could emerge (Maynard Smith and Szathmáry, 1995; Knight et al., 2001). Some suggest that the present code is a 'frozen accident' (Crick, 1968) as any further alteration of the code would imply a widespread, disastrous disruption of nearly all the proteins. Others depict scenarios that depend on primordial biochemical pathways and their evolution (Wong, 1975). Another class of models proposes that the order in the code can be explained by evolutionary selection for codes that minimize the load of translation errors and mutations (Sonneborn, 1965; Zuckerkandl and Pauling, 1965; Epstein, 1966; Goldberg and Wittes, 1966; Alff-Steinberger, 1969). Evolutionary dynamics studies in this spirit (Figureau, 1989; Ardell and Sella, 2001) demonstrated selection for optimal codes. The discovery that the genetic code is not strictly universal, that many organisms and mitochondria use variants of the standard code (Osawa et al., 1992; Knight et al., 2001), supports the suggestion that the code was not utterly frozen and remained, at least for a while, open to adaptation (Fig. 1B). This motivates us to examine a simple model in which a genetic code emerges when evolution selects for an optimal amino-acids map.

In what follows, we first discuss the code as an information channel or a map between nucleotides and amino-acids and construct an estimate for its error-load. Next, the error-load serves as the code fitness in a simplified evolutionary model. We then show that a genetic code may emerge at a transition in this information channel, from a 'non-coding' population to a 'coding' one. We then consider the topology of the map and how this topology affects the transition and the characteristics of the emergent code. We conclude by discussing, within the many limitations of the model, its general consequences for other topologies and other noisy biological codes.

## 2. A genetic code is a map with a natural fitness measure

We begin by evaluating how good a genetic code is. Without delving into the question of which came first (Maynard Smith and Szathmáry, 1995; Dyson, 1999; Gesteland et al., 1999), we assume that nucleotides and amino-acids coexist in a primitive cell. Initially, nucleotides bind nonspecifically amino-acids and might thus catalyze the polymerization of proto-proteins. We leave the exact mechanism that could, at some point, make this recognition specific completely general and focus on the evolutionary benefit this specificity offers: An organism may gain some control over protein synthesis by mapping amino-acids into nucleotide sequences in a primitive genome that can be read and translated to proteins exactly when required.

A natural measure for the reliability of such a map is the average difference between the desired amino-acid and the one that is actually read (Fig. 2). To evaluate this measure, termed *error-load*, we simply trace all the possible ways to encode an amino-acid as a nucleotide sequence in a primitive genome and decode back the amino-acid: Any amino-acid $\alpha$ is characterized by a set of chemical features that may be thought of as a point in an abstract chemical space. The amino–acid $\alpha$ is encoded by a codon $i$, as denoted by the corresponding entry in the encoder matrix $E_{\alpha i} = 1$. The codon $i$ may be read correctly as $i$ or misread as $j$ with a probability $R_{ij}$, where $R$ is the reading matrix. Finally, codon $j$ is decoded into an amino-acid $\beta$, as denoted by the relevant entry in the decoder matrix $D_{j\beta} = 1$. The original amino-acid $\alpha$ and the one that is read $\beta$ are dissimilar by some chemical distance $C_{\alpha\beta}$. Given the reading matrix $R$, the error-load $H_{ED}$ of a map specified by $E$ and $D$, is the chemical distances between original and final amino-acids $C_{\alpha\beta}$ summed over all possible paths $\alpha \rightarrow i \rightarrow j \rightarrow \beta$ weighted by the probability for each path $P_{\alpha ij\beta}$ (Fig.2),

$$(1) \quad H_{ED} = \sum_{\alpha \rightarrow i \rightarrow j \rightarrow \beta} P_{\alpha ij\beta} C_{\alpha\beta} = \sum_{\alpha,i,j,\beta} P_\alpha E_{\alpha i} R_{ij} D_{j\beta} C_{\alpha\beta},$$

where $P_\alpha$ is the probability that amino-acid $\alpha$ is required. It is apparent from Eq. 1 that the error-load depends on both the encoding of desired amino-acids (as determined by $E$) into a genome and the decoding back ($D$) into translated proteins. To achieve finer codes the encoding $E$ and the decoding $D$ must be optimized with respect to $R$.

A useful scale for the chemical distance $C_{\alpha\beta}$ is the average impact made on fitness by confusing the amino-acids $\alpha$ and $\beta$. On this scale, the error-load $H_{ED}$ is actually the overall reduction in fitness due to the imperfection of the map (Fig 2). Eq. 1 implies that the decoded amino-acids must be diverse enough to map diverse chemical properties. However, to minimize the impact of errors it is preferable to decode fewer amino-acids. Optimizing $H_{ED}$ therefore requires the balance of these two competing needs. In information theory, the error-load measure $H$ is termed *distortion* and is used to evaluate the reliability of man-made information channels (Shannon, 1960; Berger, 1971; Tlusty, 2007b). For example, live music is recorded as bits on a CD which is then read and played by a CD-player. The distortion in this case is the average difference between the original music and the one that is actually reproduced.

### 3. A steady-state model for the evolution of a genetic code

Our scenario imagines a population of competing primitive organisms. The organism with the smallest error-load takes over the population. The underlying assumption is that although a mature genetic apparatus had not yet been developed, the organisms can still reproduce and inherit, perhaps imprecisely, the mapping that their ancestors used (Dyson, 1999). We lack experimental knowledge on the environment where this evolutionary struggle took place. Still, it is plausible to assume that the population was relatively small and the selection forces are therefore weak with respect to random drift. In addition, the high noise levels in the synthesis of primitive proteins further weakens the selection. Only when modern genetic apparatus appeared did the strength of selection increase.

A family of evolutionary processes that are valid in the relevant regime of small effective population and strong drift was recently analyzed (Berg et al., 2004; Sella and Hirsh, 2005). These processes exhibit the detailed balance property: For each pair of phenotypes, *A* and *B*, the transition rates between these phenotypes, $\pi_{A \to B}$ and $\pi_{B \to A}$, and the probabilities that the population is fixed at these phenotypes, $P_A$ and $P_B$, obey the relation $P_A \times \pi_{A \to B} = P_B \times \pi_{B \to A}$, which states that the currents are equal in both directions. This makes the system completely analogous to a thermodynamic system in equilibrium (Sella and Hirsh, 2005). From this analogy it follows directly that the population reaches a steady-state where the probability of a takeover by an organism increases exponentially with its fitness *g*, $P(g) \sim \exp(g/T)$, where *T* is inversely proportional to the effective population size, $T \sim 1/N_{eff}$. Our scenario assumes that the fitness is minus the error-load $H_{ED}$. The takeover probability is therefore $P_{ED} \sim \exp(-H_{ED}/T)$. The Boltzmann-like probability $P_{ED}$ minimizes a functional analogous to the Helmholtz free energy

$$(2) \qquad F = \langle H \rangle - TS = \sum_{ED} H_{ED} P_{ED} + T \sum_{ED} P_{ED} \ln P_{ED},$$

where *<H>* is the average error-load and *S* is the entropy due to the random drift. The temperature-like parameter *T* measures the strength of random drift relative to the selection force that pushes towards fitness maximization. The smaller the population the 'hotter' it is and the easier it is for the random drift to smear the distribution over all possible codes. Larger populations are 'colder', harder to shift, and selection sharpens the distribution around codes with minimal error-loads. Although the code itself is a small system of a few dozen codons and amino-acids, the statistical mechanics approach of Eq. 2 is reasonable when the average is taken over long enough times and large enough populations. The functional *F* is used in *rate-distortion theory* to design optimal information channels (Shannon, 1960; Berger, 1971; Tlusty, 2007b). The engineer's task is to find an optimal channel (specified by the distribution $P_{ED}$) of minimal information rate (maximal entropy), whose average distortion is no more than *<H>*. The free energy (Eq. 2) is nothing else but the manifestation of this minimization with $T$ as a Lagrange multiplier. Hence, the average code described by $P_{ED}$ determines an optimized information channel.

## 4. The emergence of a genetic code as a transition in an information channel

The optimal code is described by the minimum of a 'free-energy', which brings up very naturally the possibility of describing its emergence as a transition, akin to a phase transition in statistical mechanics. Indeed, a supercritical transition is known to take place in noisy information channels (Berger, 1971; Rose et al., 1990; Graepel et al., 1997; Rose, 1998). The noisy channel is controlled by a temperature-like parameter $T$ that determines the balance between the information rate and the distortion in the same way that physical temperature controls the balance between energy and entropy. At high T the channel is totally random and it conveys zero information. At a certain critical temperature $T_c$ the information rate starts to increase continuously. One may therefore imagine a scenario in which a population proliferates and therefore its effective temperature $T \sim 1/N_{\mathit{eff}}$ decreases below $T_c$, resulting in a potential coding transition. The population may also cool down if the selection forces increase, for example as a result of an increasing complexity of the organisms.

To search for a possible transition, we follow the behavior of the steady-state takeover probability $P_{ED}$ as the population cools down. At very high temperatures or small populations, the Boltzmann distribution implies that all possible codes are equally probable. Every codon $j$ is equally likely to be translated to any of the $N_A$ amino-acids and the average decoder matrix over all possible codes is therefore $<D_{j\beta}> = \sum_{ED} P_{ED} D_{j\beta} = 1/N_A$. Similarly, every amino-acid is equally likely to be encoded by any of the $N_C$ codons, and the average encoder matrix is $<E_{\alpha i}> = 1/N_C$. We term this state *non-coding* because knowledge of the final amino-acid $\beta$ conveys no information about the original one $\alpha$ and the relation between amino-acids and codons is entirely nonspecific.

As we show below, this symmetric non-coding state ensues until the population reaches a critical temperature $T_c$, where symmetry is broken and a 'coding' state emerges. The order-parameter that indicates the symmetry-breaking is $e_{\alpha i} = <E_{\alpha i}> - 1/N_C$, where $N_C$ is the number of codons. The order parameter $e_{\alpha i}$ manifests the correlation between the amino-acids and the codons. Above $T_c$, the codons and the amino-acids are uncorrelated, $e_{\alpha i} = 0$. At this non-coding state, the encoder randomly draws a codon no matter what the amino-acid is. Thus the channel conveys no information about the amino-acids. At the transition, the symmetric state $e_{\alpha i} = 0$ is no longer a stable minimum of $F$ and an asymmetric state $e_{\alpha i} \neq 0$ takes its place. The new state may be termed *coding* because now, on average, some codons prefer to encode specific amino-acids more than others. The order-parameter $e_{\alpha i}$ measures this preference.

To locate a transition, we expand the free energy (Eq. 2) around the non-coding state $e_{\alpha i} = 0$ (Appendix A). A mean-field approximation yields the quadratic form (Rose et al., 1990; Graepel et al., 1997; Rose, 1998; Tlusty, 2007b)

$$(3) \qquad F \sim \sum_{i,j,\alpha,\beta} \left( T \delta_{ij} \delta_{\alpha\beta} - r_{ij}^2 c_{\alpha\beta} \right) e_{\alpha i} e_{\beta j}.$$

Here, the normalized misreading $r_{ij}$ is the deviation of the misreading probability $R_{ij}$ from that of a completely random reading apparatus $r_{ij} = R_{ij} - 1/N_C$, and the normalized chemical distance is $c_{\alpha\beta} = C_{\alpha\beta} - N_A^{-1} \sum_\varepsilon C_{\varepsilon\beta} - N_A^{-1} \sum_\eta C_{\alpha\eta} + N_A^{-2} \sum_{\varepsilon\eta} C_{\varepsilon\eta}$. At high temperatures, the curvature of $F$ (Eq. 3) is positive and the non-coding state is a stable minimum. However, when the temperature is decreased below a critical value $T_c$ the non-coding state is no longer a minimum of $F$. From the stability analysis of $F$ (Appendix A) it follows that the critical temperature is the product of the maximal eigenvalues, $\lambda_r$ and $\lambda_c$, of the normalized reading matrix and of the chemical distance matrix

$$(4) \qquad T_c = \lambda_r^2 \lambda_c .$$

The code appears as the mode $e_{\alpha i}$ of $F$ that corresponds to these maximal eigenvalues. An amino-acid $\alpha$ tends to be encoded by the codons $i$ for which $e_{\alpha i} > 0$. Similarly, codons where $e_{\alpha i} < 0$ will encode $\alpha$ less often. The critical temperature (Eq. 4) depends on the accuracy of the reading apparatus: if it cannot distinguish any codon then $r_{ij} = 0$ and there is no transition as $\lambda_r = 0$ and $T_c = 0$ (that is in the limit of an infinite population). $\lambda_r$ and $T_c$ increase with the accuracy of reading. This suggests another, 'isothermal' ($N_{eff}$ = const.) pathway to the transition by improving the accuracy of reading at constant population size. We note that the code-no-code transition is supercritical, that is the code emerges continuously as a small perturbation of the symmetric state. When the temperature is further decreased far below $T_c$, other modes – apart from the one that corresponds to the maximal eigenvalue – are induced and may become relevant in shaping the code. An open question is whether the code may still evolve at such lower temperatures or perhaps it freezes in the proximity of the critical temperature, in a manner similar to the slowing down of the dynamics in a glass (see point 4 in section 6 below).

**5. Topology of the codon graph and its impact on the emergent genetic code**

We have found that, within our model, the code that emerges at the transition appears as a mode $e_{\alpha i}$ that minimizes the free energy $F$ (Eq. 3). This is the mode that corresponds to the maximal eigenvalue of the normalized reading matrix $r$ (Eq. 4). In principle, knowledge of every reading probability $r_{ij}$ and every distance $c_{\alpha\beta}$ at the time of the transition would allow full derivation of $e_{\alpha i}$, the emergent code. However, as we lack this knowledge we pursue instead generic properties of the code. In particular, we look at the topology of the graph that describes the probable reading errors, with the aim of learning how it governs basic properties of the code.

To discuss the topology of errors we portray the codon space as a graph whose vertices are the codons (Fig. 3). Two codons $i$ and $j$ are linked by an edge if they are likely to be confused by misreading, that is they have a significant $R_{ij}$. We assume that two codons are most likely to be confused if all their letters except for one agree and therefore draw an edge between them. The resulting graph is natural for considering the impact of translation errors or mutations because such errors almost always involve a single letter difference, that is a movement along an edge of the graph to a neighboring vertex (Tlusty, 2007b).

The topology of a graph is characterized by its genus $\gamma$, the minimal number of holes required for a surface to embed the graph such that no two edges cross (Gross and Tucker, 1987). The more connected that a graph is the more holes are required for its minimal embedding (Fig. 3). Indeed, we find that the highly interconnected 64-codon graph is embedded in a holey, $\gamma = 41$ surface. The genus is somewhat reduced to $\gamma = 25$ if we consider only 48 effective codons (Appendix D). On the embedding surfaces, neighboring codons are those that are likely to be misread by translation errors. Our model depicts the emergence of a genetic code by the appearance a mode of the error-load function $e_{\alpha i} \neq 0$ on these surfaces. For each amino-acid $\alpha$, $e_{\alpha i}$ measures the average preference of codon $i$ to encode $\alpha$.

The free energy may have several degenerate low modes $e_{\alpha i}$, that is modes with the same minimal value of $F$. According to Courant's theorem, the first excited mode of a self-adjoint operator, such as $F$, has a single maximum (Courant and Hilbert, 1953; Davies et al., 2001). The maximum determines a single contiguous domain where a certain amino-acid is encoded (Appendix B). Thus every mode corresponds to an amino-acid and the number of modes is the number of amino-acids. This compact organization is advantageous because misreading of one codon as another codon within the same domain has no deleterious effect. For example, if the code has two amino-acids, it is evident that the error-load of an arrangement where there are two large contiguous regions, each coding for a different amino-acid, is much smaller than a 'checkerboard' arrangement of the amino-acids. Indeed, in the genetic code all amino-acids are encoded by single contiguous domains except serine that has two domains (Fig. 1A).

What can a graph's topology tell about its low energy modes? – A useful physical analogy that comes to mind is a beating drum. The elastic energy of the drum is the equivalent of the quadratic error-load free energy $F$ (Eq.3). The sound spectrum of the drum is a series of discrete modes with quantized frequencies that are known to be determined by its shape and topology (Kac, 1966). In particular, topology affects the low-frequency modes of the drum (Tlusty, 2007a). These are the smoothest, lowest elastic energy modes and are therefore the easiest to excite. In the same spirit, the evolution of the code is governed by the underlying topology of the codon graph. At the transition, only the modes with the lowest error-load emerge and these modes, just like the modes of the drum, are affected by the topology.

We suggest that the topology of the code sets an upper limit to the number of low modes, which is also the number of amino-acids. These low modes define a partition of the codon surface into domains, and in each domain a single amino-acid is encoded. The partition can optimize the average error-load (Eq. 1) by minimizing the boundaries between the domains as well as the dissimilarity between neighboring amino-acids. One may think of the partition and of the assignment of amino-acids in terms of drawing and coloring a geographical map, where each amino-acid domain is a 'country' represented by its own color. In the following, we discuss how this partition is related to another, well-known topological partition problem, the coloring problem.

In the coloring problem the goal is to find the minimal number of colors required to color an arbitrary partition of a surface such that no two bordering domains have the same color. This limit is known as the *coloring number* of the surface $chr(\gamma)$, which is determined by its topology, namely its genus $\gamma$. The coloring number of a sphere, for example, is $chr(0) = 4$ by the four-color theorem. The coloring number for any genus is determined by Heawood's formula (Ringel and Youngs, 1968)

$$(5) \qquad chr(\gamma) = \text{int}\left(\tfrac{1}{2}\left(7 + \sqrt{1 + 48\gamma}\right)\right),$$

where *int* denotes integer value. We can show that the coloring number (Eq. 5) is a bound on the number of low modes, which we suggest is the number of amino-acids.

The relation of the coloring problem to the maximal number of lowest modes may be understood by considering a mapping $\vec{x}(i)$ of the codon graph into Euclidean space (Appendix C). Each coordinate $x_\alpha(i) = u_{\alpha i}$ of this map corresponds to one of the lowest modes and the dimension of the surface is therefore equal to the number of such modes (Tlusty, 2007a). Courant's theorem dictates that the first-excited eigenmodes of a self-adjoint operator, such as $F$, divide the surface into exactly two sign-domains, in which the eigenmode is either non-negative or non-positive (Courant and Hilbert, 1953; Davies et al., 2001). It follows that the resulting surface has only one maximum and one minimum in any given direction. Surfaces with this property are termed tight. A theorem due to Banchoff (Banchoff, 1965) states that the maximal dimension of tight surfaces of genus $\gamma$, and thus the number of lowest modes, is $chr(\gamma) - 1$. Together with the additional ground-state mode, the total number of available modes, which is the maximal number of amino-acids, is the coloring number *chr*($\gamma$) (Tlusty, 2007a).

For the 64-codons graph the coloring limit is *chr*(41) = 25, while for the 48-codons graph this limit is reduced to *chr*(25) = 20 (Appendix D). Hence, a reasonable estimate for the coloring limit lies in the range of 20-25. We note that the stop codons do not change the coloring number limit, because they are segregated into a single hole next to the amino-acid tryptophan (Fig. 1A) and such a hole is known to keep the number of modes unchanged (Banchoff and Kuhnel, 1997). This stop hole serves effectively as a boundary of the embedding surface, because no amino-acid is chemically similar to a termination signal.

The present approach also suggests a pathway for the evolution of the present-day code from simpler codes, driven by the increasing accuracy of improving translation machinery. Early translation machinery corresponds to smaller graphs since indiscernible codons are described by the same vertex. As the accuracy improves these codons become discernible and the corresponding vertex splits. This gives rise to a larger graph that can accommodate more amino-acids. Table 1 shows the predicted number of amino-acids for various code topologies. Early translation machinery that could discern, for instance, only three letters in the first two positions of the codon could read nine codons and encode a maximum of seven amino-acids (because seven is the coloring number of the torus – Fig. 2B). As the machinery becomes more stringent and four letters are discernable, still in the first two positions, the number of amino-acids increases to eleven. This is in accord with the number of amino-acids observed (around 10-12) in simulations of a population with a genetic code of four-letter doublets in the regime of high error tolerance that corresponds to our transition (Ardell and Sella, 2001). Finally, the present-day translation machinery with a four letter code and 48-64 codons (no discrimination between U and C in the third position) gave rise to 20-25 amino-acids. One may think of future improvement that will remove the ambiguity in the third position (64 discernable codons). This is predicted to enable a stable expansion of the code up to 25 amino-acids.

## 6. Limitations of the model and potential experimental tests

The present model exhibits weaknesses and leaves several open questions:

1. The model takes for granted that there is an assignment of nucleotide triplets to amino-acids. It does not explain how exactly such assignment originated in the first place nor why specifically these amino-acids were chosen and not others. The model does not answer any of these fundamental questions. Instead, it focuses on the evolutionary mechanism that might drive the emergence of a genetic code and explain its generic properties. However, the model does support the possibility that when new amino-acids were introduced, they were chosen by some similarity criteria, such as chemical similarity or the precursor-product kinship suggested by the coevolution theory (Wong, 1975).

2. We used a specific class of evolutionary dynamics that exhibit a quasi steady-state behavior. We employed these dynamics for the sake of simplicity and to place our model within the framework of equilibrium statistical mechanics, where the notion of transition is brought up very naturally. An underlying assumption is that the genetic code emerges in a relatively small population under the influence of weak selection forces. The occurrence of transition needs to be tested in other models of evolutionary dynamics.

3. The coloring number is an upper limit to the number of amino-acids, but there still remains the question of why the code should exhaust this maximal number. One possible mechanism that would drive the code to the maximal diversity limit is evolution itself. The misreading rates $R_{ij}$ may coevolve with the code and may be selected to enable maximal amino-acid diversity. The model, however, does not suggest a concrete mechanism for this coevolution.

4. The model assumes that the code was shaped around the time of the transition. An underlying assumption is that not long after the transition it became too costly to alter the code because any further change would be devastating due to the global impact on the proteins. In that sense, our scenario is of 'almost-frozen' dynamics (Crick, 1968). What might have frozen the dynamics are the increased selection forces after the emergence of the code. As organisms became more complex a change in the code implied an unbearable modification of too many proteins. As a result, the only relevant modes are the smooth ones that appear at the coding transition.

5. The wobble rules assume that only 48 codons are distinguishable due to physiochemical limitations of the translation apparatus (Crick, 1966; Osawa et al., 1992) and the resulting codon graph leads to a sharp limit of 20 amino-acids. However, if the wobble ambiguity is removed there are 64 discernible codons and the maximal number of amino-acids increases to 25 (Table 1). Although 20 and 25 are not too far apart, it is still unclear why evolution froze before it could improve the translation apparatus so it can discern all the 64 codons.

6. The amino-acid serine occupies two isolated domains, if only single-letter edges are considered. It might be that the code froze in this slightly suboptimal configuration, but the present form of the model is too idealized and simplified to account for such a subtle effect.

7. The most probable translation errors at the time the code was determined are assumed to involve only a single letter difference. This hypothesis may be wrong due to chemical details of the reading machinery, which might allow certain two- and three-letter errors. Still, the model can treat other error patterns as long as their topology is given (Table 1).

Many of the listed weaknesses reflect the substantial obstacle we face — the absence of experimental knowledge about the chemical environment where the code emerged. One major theme of the present work is the examination of a possibility that the code emerged during an evolutionary process analogous to a transition in an information channel. The depicted scenario is rather generic and phenomenological, and the occurrence of transition does not depend finely on the missing details of the evolutionary pathway.

The second theme of the model is that although not much is known about the primordial environment, minimal assumptions about the topology of probable errors can yield characteristics of possible genetic codes, in particular limits on the number of encoded amino-acids.

As we lack sound theoretical grounds for estimating the model parameters, especially the misreading rates in early evolution, experimental studies would be most valuable. One direction would be tracking the evolution of organisms whose code was altered to include more or less amino-acids than the optimal number. Recent work (Wang et al., 2001; Chin et al., 2003) demonstrated alteration of the code by introducing a modified tRNA and a matching aminoacyl-tRNA synthetase that charged the tRNA with a nonstandard amino-acid. In principle, one might be able to observe expansion or reduction of an altered code to the optimal number or at least some precursor of such change. The transition between codes may be softened by keeping the original tRNA and smoothly changing the expression level of the competing tRNAs. Also, reassignment of a relatively rarely used codon would minimize the impact on fitness.

As the present approach applies for any generic noisy code, another direction, perhaps more accessible, would be the study of other biological and artificial coding systems (Tlusty, 2007b). Recognition processes may be thought of as codes since they specifically associate molecules, for example protein-DNA interactions map proteins to their respective DNA binding sites. One possibility is treating transcription networks in terms of an error-prone code (Itzkovitz et al., 2006; Shinar et al., 2006) In this case, one may construct a graph whose vertices are the DNA binding sites that are connected by edges if they are likely to be confused. The code is the assignment of transcription factors to binding sites that are the 'codons' in this case. Similar topological arguments may yield limits to the number of transcription factors and their organization in the network.

**Tables**

| Code | 1st position | 2nd position | 3rd position | Number of codons, $V$ | $E$ | $F$ | $\gamma$ | Maximal number of amino-acids, $chr(\gamma)$ |
|---|---|---|---|---|---|---|---|---|
| 4-base singlets | 1 | 4 | 1 | 4 | 6 | 4 | 0 | **4** |
| 3-base doublets | 3 | 3 | 1 | 9 | 18 | 9 | 1 | **7** |
| 4-base doublets | 4 | 4 | 1 | 16 | 48 | 24 | 5 | **11** |
| 16 codons | 4 | 4 | 2 | 32 | 112 | 56 | 13 | **16** |
| 48 codons | 4 | 4 | 3 | 48 | 192 | 96 | 25 | **20** |
| 4-base triplets | 4 | 4 | 4 | 64 | 288 | 144 | 41 | **25** |

Table 1. **Topological limit to the number of amino-acids.** The codes are specified by the number of discernable letters in the three positions of the codons. The number of codons ($V$), edges ($E$) and faces ($F$) determine the genus ($\gamma$). The topological limit on the maximal number of amino-acids is the coloring number, $chr(\gamma)$ (Eq. 5).

**Figure Legends**

Fig 1. **The standard genetic code and its variants**. **(A)** The 64 triplet codons are listed with the amino-acids or the stop signal (TER) they encode. The codons for a given amino-acids are grouped into contiguous domains in the table. **(B)** Arrows point from codons to amino-acids they encode in variant codes (Data from (Knight et al., 2001) and http://www.ncbi.nlm.nih.gov/Taxonomy/Utils/wprintgc.cgi?mode=c). All the variants use the same twenty amino-acids, suggesting that the number twenty is a universal invariant. Most of the variants involve one-letter changes (dark arrows) while some involve two-letter changes (light arrows), probably obtained via two consecutive one-letter changes. This suggests that the evolution of the code may behave like a smooth continuous flow in the codon space (Eq. 3).

Fig. 2. **The genetic code as a map or an information channel**
$N_A$ amino-acids reside in an abstract 'chemistry' space whose points are all possible combinations of chemical properties. The $N_C$ codons, say the 64 triplets, reside in a space of their own. The probability that amino-acid $\alpha$ is required is $P_\alpha$. An amino–acid $\alpha$ is mapped to codon $i$, as denoted by the binary matrix $E_{\alpha i} = 1$ (for all other codons $k \neq i$, $E_{\alpha k} = 0$ and $\sum_j E_{\alpha j} = 1$). The codon $i$ may be read correctly as $i$ or misread as $j$ with a probability $R_{ij}$ (the sum over all reading probabilities obeys $\sum_i R_{ij} = 1$). Finally, codon $j$ is translated back to an amino-acid $\beta$, as denoted by the binary matrix $D_{j\beta}$ (the sum over all reading probabilities obeys $\sum_\beta D_{j\beta} = 1$). Thus, the probability of each path $\alpha \rightarrow i \rightarrow j \rightarrow \beta$ is $P_\alpha E_{\alpha i} R_{ij} D_{j\beta}$. The matrices $E$ and $D$ specify the code and the reading apparatus is characterized by the misreading matrix $R$. The chemical dissimilarity between the original amino-acid and the final one is $C_{\alpha\beta}$. The error-load of a given map $H_{ED}$ is the weighted average of chemical distances between original and final amino-acids $C_{\alpha\beta}$ over all possible paths (Eq.1). The error load is actually the distortion of the information channel described by the map.

Fig. 3. **Topology of the codon graph.**

To illustrate the topology of the natural codon graph we consider two simpler codes. **(A)** The first is a triplet code of a two-letter alphabet, say U and C. The vertices are triplets of a two-letter alphabet and edges link codons differing by one letter. The resulting code graph takes the form of a sphere. **(B)** The second example is a doublet code of a three-letter alphabet, U, C, and A. The resulting graph exhibits a toroidal topology. The different topologies of the two graphs are manifested by their genus $\gamma$, the minimal number of holes required for a surface to embed the graph such that no two edges cross (Gross and Tucker, 1987). In the sphere, each vertex has three neighbors and $\gamma = 0$, whereas each vertex of the torus has four neighbors and $\gamma = 1$. The coloring number of the sphere is $chr(0)=4$ by the four-color theorem, while the torus has $chr(1) = 7$ (Eq. 5) **(C)** The 48-codon graph is a triplet code of a four-letter alphabet, where at the third position only three letters can be discerned: While the genus of the two simple examples is apparent by inspection, to find the genus of the natural code we need to employ Euler's formula $\gamma = 1 - \frac{1}{2}(V - E + F)$, where $V$ is the number of vertices of the graph, $E$ – the number of its edges and $F$ is the number of faces of the surface embedding the graph. In the torus, for instance, there are $V = 9$ vertices, each of which has $d = 4$ neighbors. It follows that there are $E = V \times (d/2) = 18$ edges and $F = V \times (d/4) = 9$ quadrilateral faces, yielding a genus $\gamma = 1$. Similarly, in the natural code, where each of the $V = 48$ vertices has $d = 8$ neighbors, there are $E = 192$ edges and $F = 96$ quadrilateral faces, leading to a holey, $\gamma = 25$ embedding surface (Appendix C). The corresponding coloring number is $chr(25) = 20$. The figure shows a portion of the surface around the codon *AAA* with its 8 neighboring codons and 8 adjacent faces.

|   | U | C | A | G |
|---|---|---|---|---|
| U | UUU Phe<br>UUC Phe<br>UUA Leu<br>UUG Leu | UCU Ser<br>UCC Ser<br>UCA Ser<br>UCG Ser | UAU Tyr<br>UAC Tyr<br>UAA TER<br>UAG TER | UGU Cys<br>UGC Cys<br>UGA TER<br>UGG Trp |
| C | CUU Leu<br>CUC Leu<br>CUA Leu<br>CUG Leu | CCU Pro<br>CCC Pro<br>CCA Pro<br>CCG Pro | CAU His<br>CAC His<br>CAA Gln<br>CAG Gln | CGU Arg<br>CGC Arg<br>CGA Arg<br>CGG Arg |
| A | AUU Ile<br>AUC Ile<br>AUA Ile<br>AUG Met | ACU Thr<br>ACC Thr<br>ACA Thr<br>ACG Thr | AAU Asn<br>AAC Asn<br>AAA Lys<br>AAG Lys | AGU Ser<br>AGC Ser<br>AGA Arg<br>AGG Arg |
| G | GUU Val<br>GUC Val<br>GUA Val<br>GUG Val | GCU Ala<br>GCC Ala<br>GCA Ala<br>GCG Ala | GAU Asp<br>GAC Asp<br>GAA Glu<br>GAG Glu | GGU Gly<br>GGC Gly<br>GGA Gly<br>GGG Gly |

**Fig. 1A**

|   | U | C | A | G |
|---|---|---|---|---|
| U | UUU Phe<br>UUC Phe<br>UUA Leu<br>UUG Leu | UCU Ser<br>UCC Ser<br>UCA Ser<br>UCG Ser | UAU Tyr<br>UAC Tyr<br>UAA TER<br>UAG TER | UGU Cys<br>UGC Cys<br>UGA TER<br>UGG Trp |
| C | CUU Leu<br>CUC Leu<br>CUA Leu<br>CUG Leu | CCU Pro<br>CCC Pro<br>CCA Pro<br>CCG Pro | CAU His<br>CAC His<br>CAA Gln<br>CAG Gln | CGU Arg<br>CGC Arg<br>CGA Arg<br>CGG Arg |
| A | AUU Ile<br>AUC Ile<br>AUA Ile<br>AUG Met | ACU Thr<br>ACC Thr<br>ACA Thr<br>ACG Thr | AAU Asn<br>AAC Asn<br>AAA Lys<br>AAG Lys | AGU Ser<br>AGC Ser<br>AGA Arg<br>AGG Arg |
| G | GUU Val<br>GUC Val<br>GUA Val<br>GUG Val | GCU Ala<br>GCC Ala<br>GCA Ala<br>GCG Ala | GAU Asp<br>GAC Asp<br>GAA Glu<br>GAG Glu | GGU Gly<br>GGC Gly<br>GGA Gly<br>GGG Gly |

**Fig . 1B**

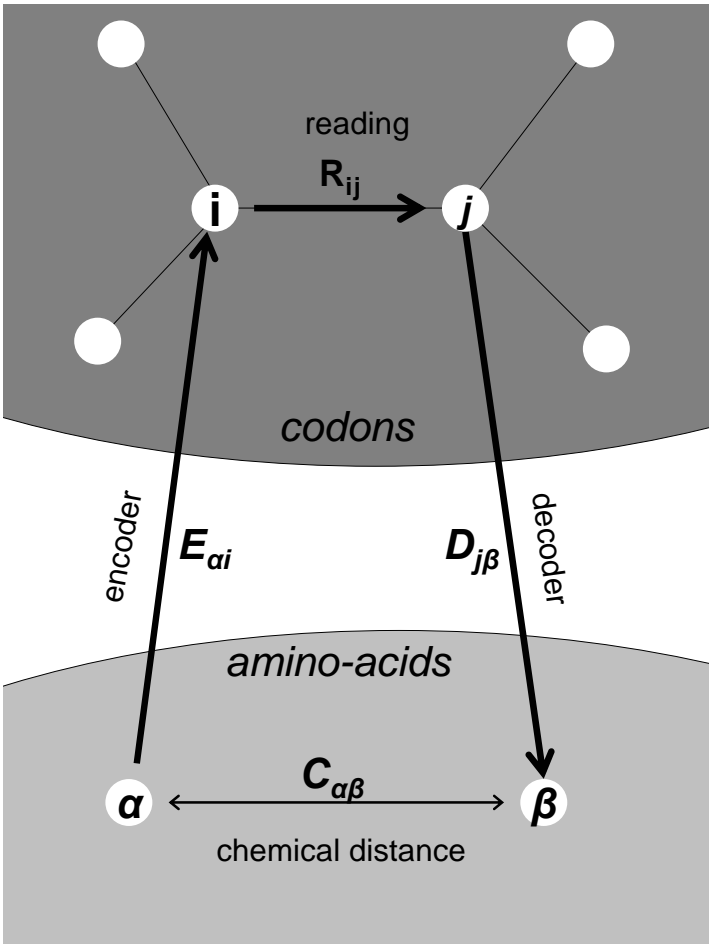

**Fig. 2**

**A**

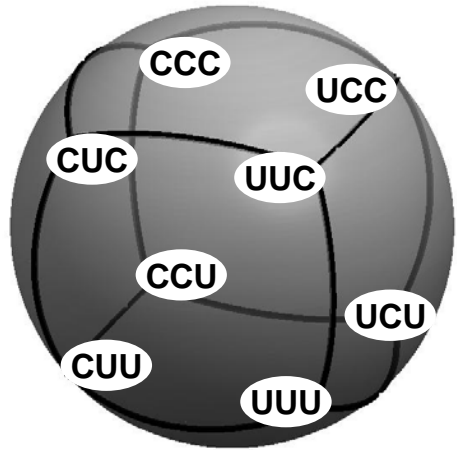

**B**

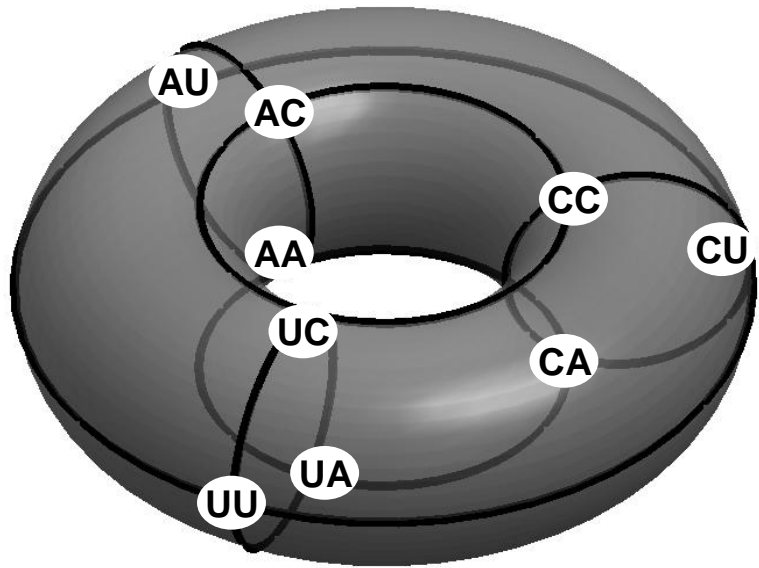

**Fig. 3**

**C**

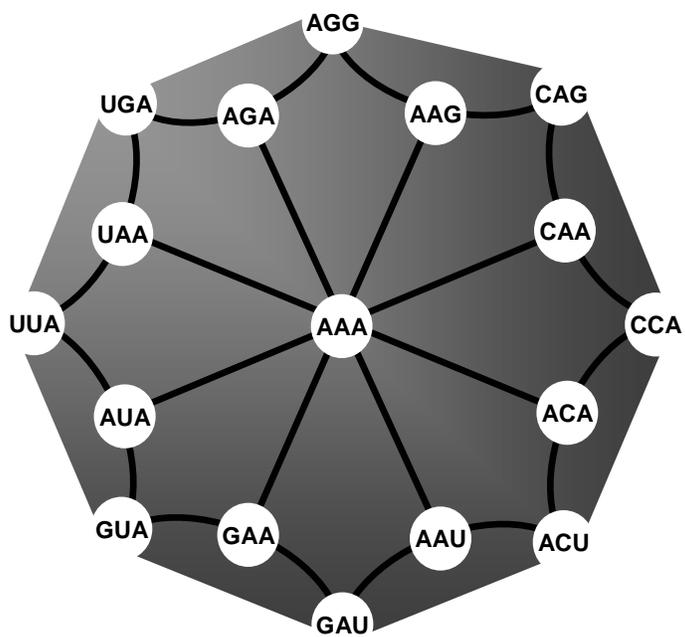

**Fig. 3**

**Appendix A**

*A mean-field model for the transition* – The equivalence between the problem of optimizing an information channel and equilibrium statistical mechanics (Shannon, 1960; Berger, 1971) was used to study transitions in clustering problems (Rose et al., 1990). The quadratic expansion of an effective free energy and the consequent critical phase transition were derived in studies of self-organized maps and classification by methods of deterministic annealing (Graepel et al., 1997; Graepel et al., 1998; Rose, 1998). In the following, we adapt the deterministic annealing formalism for the coding problem. We describe the use of statistical mechanics methods to obtain a mean-field approximation for the error-load free energy $F$ (Eq.2). We then expand this mean-field expression around the 'non-coding' symmetric state to obtain the critical modes at the transition (Eqs. 3-4).

To begin with, each map is specified by the two binary matrices $E$ and $D$. The encoder $E$ is an $N_A \times N_C$ binary matrix and the decoder $D$ is an $N_C \times N_A$ binary matrix, where $N_A$ and $N_C$ are the number of amino-acids and codons, respectively (Fig. 2). The encoder is subject to the conservation relation $\sum_j E_{\alpha j} = 1$ that assures that each amino-acid is encoded by one and only one codon. Similarly, $\sum_\beta D_{j\beta} = 1$ assures that every codon is translated into one and only one amino-acid. In terms of statistical mechanics, $E$ and $D$ determine a 'microstate', that is the instantaneous map used by the population. As we show below, $E$ and $D$ are actually related and thus one of the matrices, say $E$, is enough to specify the map. The thermodynamic average or the 'macro-state' is described by the order parameter $e_{\alpha i}$, which is the deviation of the averaged $E$ from the symmetric state $e_{\alpha i} = \langle E_{\alpha i}\rangle - 1/N_C$. A transition occurs when this macro-state ceases to be symmetric, i.e. $e_{\alpha i} \neq 0$. The error-load of each map $H_{ED}$ is given by the sum (Eq. 1, Fig. 2)

(A1) $$H_{ED} = \sum_{\alpha \to i \to j \to \beta} P_{\alpha i j \beta} C_{\alpha\beta} = \sum_{\alpha,i,j,\beta} P_\alpha E_{\alpha i} R_{ij} D_{j\beta} C_{\alpha\beta}$$

Minimization of the free energy $F$ (Eq.2) yields the familiar Boltzmann distribution for the probability of each map

(A2) $$P_{ED} \sim \exp(-H_{ED}/T),$$

and the partition function for the free energy

(A3) $$F = -T\log \sum_{ED} P_{ED} = -T\log \sum_{ED} \exp(-H_{ED}/T).$$

The encoder $E$ and the decoder $D$ are related through Bayes' theorem

(A4) $$P_j D_{j\beta} = P_\beta \sum_i E_{\beta i} R_{ij},$$

where $P_j$ is the marginal probability of codon $j$. $P_j$ can be expressed as a sum over all possible ways to encode an-amino acid as codon $k$ and misread it as $j$

(A5) $$P_j = \sum_{\gamma,k} P_\gamma E_{\gamma k} R_{kj}.$$

For technical convenience, we account for the different probabilities by sampling each amino-acid a number of times that is proportional to its actual probability. With this convention, Eq. A4 becomes $D_{j\beta} = (1/P_j)\sum_i E_{\beta i} W_{ij}$ with $P_j = \sum_{\gamma,k} E_{\gamma k} W_{kj}$ from Eq. A5. By substitution into A1 the error-load takes the form

(A6) $$H_{ED} = \sum_{\alpha,i,j,k,\beta} \frac{E_{\alpha i} E_{\beta k} R_{ij} R_{kj} C_{\alpha\beta}}{\sum_{\gamma,m} E_{\gamma m} R_{mj}}.$$

The error-load (Eq. A6) is non-linear and calculation of its corresponding partition function and free energy (Eq. A3) would be rather hard. Instead, following the usual statistical mechanics procedure, we approximate the error-load (Eq. A6) by the linear mean-field expression

(A7) $$H_0 = \sum_{\alpha,i} E_{\alpha i} \Psi_{\alpha i},$$

where $\Psi_{\alpha i}$ are the effective fields that are found below by variational minimization. Within this mean-field approximation, the free energy (S3) becomes

(A8) $$F_0 = -T \log \sum_E \exp\left(-\frac{1}{T}\sum_{\alpha,i} E_{\alpha i}\Psi_{\alpha i}\right) = -T\sum_\alpha \log \sum_i \exp(-\Psi_{\alpha i}/T).$$

The last equality was obtained using the fact that for each amino-acid $\alpha$ exactly one of the $E_{\alpha i} = 1$ and all the rest are zero. An upper limit to the actual free energy is given by the mean-field variational theorem

(A9) $$F \leq F_{\text{eff}} = F_0 + \langle H_{ED} - H_0 \rangle,$$

where the average is with respect to the mean-field distribution $P \sim \exp(-H_0/T)$. The effective fields $\Psi_{\alpha i}$ are found by minimizing $F_{\text{eff}}$ (Eq. S9) with respect to $\Psi_{\alpha i}$

(A10) $$\Psi_{\alpha i} = \sum_{j,\beta} R_{ij} \langle D_{j\beta} \rangle \left(2 C_{\alpha\beta} - \sum_\gamma \langle D_{j\gamma} \rangle C_{\beta\gamma}\right),$$

where the average decoder is determined by Eqs. (A4-A5)

$$\text{(A11)} \qquad \langle D_{j\beta}\rangle = \frac{\sum_i \langle E_{\beta i}\rangle R_{ij}}{\sum_{\alpha,i}\langle E_{\alpha i}\rangle R_{ij}}.$$

We find from Eq. A8 that the average encoder is given by the Boltzmann distribution

$$\text{(A12)} \qquad \langle E_{\alpha i}\rangle = \frac{\partial F_0}{\partial \Psi_{\alpha i}} = \frac{\exp(-\Psi_{\alpha i}/T)}{\sum_j \exp(-\Psi_{\alpha j}/T)}.$$

The mean-field free energy $F_{\text{eff}}$ as a function of the average encoder and temperature is found by substitution of Eqs. A6-A8, A10-A12 into A9.

Next, we locate the transition and the emergence of a coding state at the critical temperature. At high temperatures, or small effective populations, the non-coding state is symmetrical $\langle E_{\alpha i}\rangle = 1/N_C$. A transition occurs if the free energy $F_{\text{eff}}$ becomes unstable with respect to small fluctuations around the 'non-coding' state. The fluctuations are described by the order-parameter $e_{\alpha i} = \langle E_{\alpha i}\rangle - 1/N_C$ and the stability is determined by examining the quadratic Taylor expansion of the free energy with respect to $e_{\alpha i}$

$$\text{(A13)} \qquad F \simeq F_{\text{eff}}\big|_{e_{\alpha i}=0} + \frac{N_C}{N_A}\sum_{i,j,\alpha,\beta}\left(T\delta_{ij}\delta_{\alpha\beta} - r_{ij}^2 c_{\alpha\beta}\right)e_{\alpha i}e_{\beta j},$$

where $r^2_{ij} = \sum_k r_{ik}r_{kj}$ is the square of the normalized misreading matrix $r_{ij} = R_{ij} - 1/N_C$ and $\delta$ is the Kronecker symbol. In Eq.3 we omitted the irrelevant constant term and the factor $N_C/N_A$. In Eq. A13, $c_{\alpha\beta}$ is the normalized chemical distance

$$\text{(A14)} \qquad c_{\beta\gamma} = C_{\beta\gamma} - N_A^{-1}\sum_\varepsilon C_{\varepsilon\gamma} - N_A^{-1}\sum_\eta C_{\beta\eta} + N_A^{-2}\sum_{\varepsilon,\eta} C_{\varepsilon\eta}.$$

At high temperature it is clear that the quadratic form is positive-definite for any fluctuation $e_{\alpha i}$. We note that in Eq. A13 the amino-acid space and the codon space are decoupled and the quadratic term is actually a tensor product

$$\text{(A15)} \qquad \sum_{i,j,\alpha,\beta}\left(T\delta_{ij}\delta_{\alpha\beta} - r_{ij}^2 c_{\alpha\beta}\right)e_{\alpha i}e_{\beta j} = e^t\left(T I_C \otimes I_A - r^2 \otimes c\right)e,$$

where $r$ denotes the $N_C \times N_C$ matrix $r_{ij}$, $c$ denotes the $N_A \times N_A$ matrix of chemical distances $c_{\alpha\beta}$ (Eq. A14). $I_C$ and $I_A$ are the unit matrices in the codon and amino-acid spaces, respectively. The fluctuation $e_{\alpha i}$ is written as the vector $e$ of length $N_C \times N_A$.

It follows immediately from the tensor product (Eq. A15) that the quadratic form ceases to be positive-definite at a critical temperature that is the product of the maximal eigenvalues of $c$ and $r^2$ (Eq. 4)

(A16) $$T_c = \lambda_{r^2} \lambda_c = \lambda_r^2 \lambda_c.$$

Eq. A15 also implies that the unstable eigenmode $e_{\alpha i}$ that describes the emergent code is also a tensor product of the corresponding eigenmodes of $c$ and $r^2$

(A17) $$u = u_{r^2} \otimes u_c.$$

In the next section we discuss the generic properties of the unstable modes and their relation to the topology of the misreading matrix $r$.

**Appendix B**

*Courant Theorem, tight embedding and the coloring problem* – It follows from the quadratic expansion around the symmetric state (Eqs. 3 and A15) that the unstable mode is an eigenmode of the squared normalized misreading matrix $r^2$. The stochastic process of misreading may be thought of as diffusion on the codon graph (Kac, 1966). Formally speaking, it implies that the misreading matrix $R$ may be rewritten in terms of the discrete Laplacian operator, $\Delta$, defined by

(B1) $$\Delta_{ij} = \begin{cases} -R_{ij} & : i \neq j \\ \sum_{j \neq i} R_{ij} & : i = j \end{cases}.$$

Thus, $\Delta$ is a $N_C \times N_C$ matrix, where $N_C$ is the number of codons. For a completely random reading apparatus that cannot distinguish any codon the misreading is $R_C = 1/N_C$. It follows from Eq. B1 that the normalized misreading $r = R - 1/N_C$ is actually the difference between the codon graph Laplacian $\Delta$ and the Laplacian of the completely random reader $\Delta_C = I - 1/N_C$

(B2) $$r = -(\Delta - \Delta_C).$$

The unstable mode corresponds to the maximal eigenvalue of $r$. Hence, from Eq. B2 it follows that the modes with a maximal eigenvalue $\lambda_r$ correspond to the small eigenvalues of $\Delta$. The minimal eigenvalue of any Laplacian is 0 and the corresponding eigenmode is uniform (the Laplacian of a uniform mode vanishes because it measures the average difference between a codon and its neighbors). The eigenvalues of $\Delta_C$ are 0 (single) and 1 (multiplicity n-1). It follows that any mode that is orthogonal to the minimal uniform mode is also a mode of $\Delta_C$ with an eigenvalue 1. Consequently, the mode that corresponds to the maximal eigenvalue of $r$, $\lambda_r$, is the second, or first-excited mode, of the Laplacian $\Delta$ with an eigenvalue $\lambda_\Delta$

(B3) $$\lambda_\Delta = 1 - \lambda_r.$$

The second eigenmodes of the Laplacian are the smoothest of the non-uniform modes. These modes divide the codon graph into domains with minimal boundaries where the amino-acids reside. It is apparent that these modes will exhibit minimal error load and are therefore the first modes to be destabilized at the transition. The critical temperature (Eqs. 4 and A16) is related to the second eigenvalue of the Laplacian by

(B4) $$T_c = \lambda_r^2 \lambda_c = (1 - \lambda_\Delta)^2 \lambda_c.$$

According to Courant's theorem and its discrete analogue (Davies et al., 2001), the first excited modes of a self-adjoint operator such as $\Delta$ have exactly one maximum and one minimum (two nodal domains). To find how this geometric property sets a limit on the multiplicity $m$ of the first excited modes we consider a map of the graph into $\mathbb{R}^m$ defined by a vector whose components are the $m$ first-excited modes (Tlusty, 2007a)

(B5) $$\vec{x}(i) = (e_{1i}, e_{2i}....e_{mi}).$$

Each coordinate $x_\alpha(i) = e_{\alpha i}$ of this map corresponds to one of the lowest modes and the dimension of the map is therefore the number of modes. Since any linear combination of the first-excited mode $\sum_\alpha n_\alpha e_{\alpha j}$ is also a first-excited mode of the Laplacian, the surface defined by this mapping has only one maximum and one minimum in any given direction. Surfaces of this type are termed *tight*, and have the 'two-piece property': No plane can cut them into more than two pieces, a generalization of convexity to non-zero genus surfaces. Banchoff (Banchoff, 1965) proved that tight *m*-dimensional polyhedral surfaces like the embedding surface of the codon graph contain the complete graph $K_{m+1}$ (as a set). The map-coloring theorem (Ringel and Youngs, 1968) implies that the maximal number of vertices of a complete graph contained in a surface is equal to the coloring number $chr(\gamma)$ of that surface (Banchoff and Kuhnel, 1997). The limit on the maximal dimension of the map $x(i)$, which is the number of first-excited modes, therefore satisfies $m \leq chr(\gamma) - 1$. This limit was conjectured for smooth surfaces (Colin de Verdiere, 1986). Together with the ground-state mode the total number of available modes, which is the number of amino-acids, is the coloring number *chr(γ)*.

**Appendix C**

*Graph embedding* – The codons are all possible words of length *l* (*l* = 3 for the genetic code) that can be made of an *N*-letter alphabet (*N* = 4 for the genetic code). The codons are represented as vertices in a graph whose edges connect codons that differ by one letter and are therefore likely to misread or mutate to each other. Each of the *l* positions in the codon corresponds to a complete graph $K_N$ (A graph in which all pairs of the *N* vertices are connected). The overall graph is the graph-product of the *l* complete graphs. Doublets of three-letter alphabet, for instance, form a torus which is the product of two triangles, $K_3 \times K_3$. Similarly, the natural code is the product of two tetrahedra for the first two positions and a triangle for the third position $K_4 \times K_4 \times K_3$. The genus of a graph is the minimal genus of a surface in which one can embed the graph with no crossing edges (Gross and Tucker, 1987). Embeddings of these type of product graphs (Table 1) were obtained through analysis of the graph's automorphism group by methods used in the theory of regular maps (Conder and Dobcsanyi, 2001). Given the vertices and edges, the embedding was determined by enumerating all the faces (Gross and Tucker, 1987). In particular, the embedding of the natural code graph $K_4 \times K_4 \times K_3$ is described by its the 96 quadrilateral faces (listed in the next section). The embedding surface has a genus $\gamma = 25$. We note that in this case, at least one of the 25 handles is twisted in the form of a Klein bottle since the surface is non-orientable (Here, Euler's genus is defined as $\gamma = 1 - \frac{1}{2}(V - E + F)$, independent of orientability).

**Appendix D**

*An embedding of the 48 codon graph* – The table lists the *F* = 96 quadrilateral faces of the embedding of the codon graph $K_4 \times K_4 \times K_3$ calculated by group symmetry techniques (Conder and Dobcsanyi, 2001). The codons of each face are listed in a cyclic order where edges connect consecutive codons. U or C in the third position of the codon are indiscernible and denoted by U. The codon graph has therefore *V* = 48 vertices and *E* =

192 edges. Hence, Euler's characteristic is $\chi = V - E + F = 48 - 192 + 96 = -48$ and the corresponding genus (actually half of the cross-cap number) is $\gamma = 1 - \frac{1}{2}\chi = 25$. The coloring number is determined by Eq. 5, $chr(25) = 20$. We note that Banchoff's theorem (Banchoff, 1965) and Heawood's formula (Ringel and Youngs, 1968) (Eq. 5) are valid also for non-orientable embeddings such as the one described by the table below.

| | | | |
|---|---|---|---|
| AAA AAG AGG AGA | AGG AGU ACU ACG | GAG GAU GUU GUG | UAA UAG UGG UGA |
| AAA AAG CAG CAA | AGG AGU UGU UGG | GAG GAU CAU CAG | UAA UAU UCU UCA |
| AAA AAU ACU ACA | AGG AUG GUG GGG | GAG GGG CGG CAG | UAA UAU CAU CAA |
| AAA AAU GAU GAA | AGG AUG CUG CGG | GAG GCG UCG UAG | UAA UUA CUA CAA |
| AAA AGA UGA UAA | AGG ACG GCG GGG | GAU GGU UGU UAU | UAG UAU UUU UUG |
| AAA AUA GUA GAA | AGU AUU CUU CGU | GAU GGU CGU CAU | UAG UUG CUG CAG |
| AAA AUA UUA UAA | AGU ACU GCU GGU | GAU GCU UCU UAU | UAG UCG CCG CAG |
| AAA ACA CCA CAA | AUA AUG ACG ACA | GGA GGG UGG UGA | UAU UUU CUU CAU |
| AAG AAU AUU AUG | AUA AUG CUG CUA | GGA GGU GUU GUA | UGA UGU UUU UUA |
| AAG AAU UAU UAG | AUA AUU GUU GUA | GGA GUA UUA UGA | UGA UGU CGU CGA |
| AAG AGG UGG UAG | AUA ACA UCA UUA | GGA GCA CCA CGA | UGA UCA CCA CGA |
| AAG AUG GUG GAG | AUG AUU UUU UUG | GGG GGU GCU GCG | UGG UGU UCU UCG |
| AAG ACG GCG GAG | AUG ACG UCG UUG | GGG GGU CGU CGG | UGG UUG CUG CGG |
| AAG ACG CCG CAG | AUU ACU UCU UUU | GGG GUG UUG UGG | UGG UCG CCG CGG |
| AAU AGU UGU UAU | AUU ACU CCU CUU | GGU GUU UUU UGU | UGU UCU CCU CGU |
| AAU AGU CGU CAU | ACA ACG CCG CCA | GUA GUG GCG GCA | UUA UUG UCG UCA |
| AAU AUU GUU GAU | ACA ACU GCU GCA | GUA GUG UUG UUA | UUA UUU CUU CUA |
| AAU ACU CCU CAU | ACG ACU UCU UCG | GUA GCA CCA CUA | UCA UCU CCU CCA |
| AGA AGG CGG CGA | GAA GAG GGG GGA | GUG GUU CUU CUG | CAA CAG CGG CGA |
| AGA AGU AUU AUA | GAA GAG UAG UAA | GUG GCG CCG CUG | CAA CAU CCU CCA |
| AGA AGU GGU GGA | GAA GAU GCU GCA | GUU GCU UCU UUU | CAG CAU CUU CUG |
| AGA AUA CUA CGA | GAA GGA CGA CAA | GUU GCU CCU CUU | CGA CGU CUU CUA |
| AGA ACA GCA GGA | GAA GUA CUA CAA | GCA GCG UCG UCA | CGG CGU CCU CCG |
| AGA ACA UCA UGA | GAA GCA UCA UAA | GCG GCU CCU CCG | CUA CUG CCG CCA |

30